\documentclass[preprints,article,accept,moreauthors,pdftex]{mdpi}

\setlist[itemize]{leftmargin=*,labelsep=5.8mm}
\setlist[enumerate]{leftmargin=*,labelsep=4.9mm}

\usepackage{longtable}

\firstpage{1}
\pubvolume{xx}
\issuenum{1}
\articlenumber{5}
\pubyear{2019}
\copyrightyear{2019}
\history{Received: date; Accepted: date; Published: date}
\updates{yes} 

\Title{DeepTriangle: A Deep Learning Approach to Loss Reserving}

\Author{Kevin Kuo$^{1, *}$\href{https://orcid.org/0000-0001-7803-7901}{\orcidicon}}

\AuthorNames{Kevin Kuo}

\address{%
$^{1}$ \quad Kasa AI; \href{mailto:kevin@kasa.ai}{\nolinkurl{kevin@kasa.ai}}\\
}
\corres{Correspondence: \href{mailto:kevin@kasa.ai}{\nolinkurl{kevin@kasa.ai}}}

\abstract{We propose a novel approach for loss reserving based on deep neural networks. The approach allows for joint modeling of paid losses and claims outstanding, and incorporation of heterogeneous inputs. We validate the models on loss reserving data across lines of business, and show that they improve on the predictive accuracy of existing stochastic methods. The models require minimal feature engineering and expert input, and can be automated to produce forecasts more frequently than manual workflows.}


\begin{document}

\hypertarget{introduction}{%
\section{Introduction}\label{introduction}}

In the loss reserving exercise for property and casualty insurers, actuaries are concerned with forecasting future payments due to claims. Accurately estimating these payments is important from the perspectives of various stakeholders in the insurance industry. For the management of the insurer, the estimates of unpaid claims inform decisions in underwriting, pricing, and strategy. For the investors, loss reserves, and transactions related to them, are essential components in the balance sheet and income statement of the insurer. And, for the regulators, accurate loss reserves are needed to appropriately understand the financial soundness of the insurer.

There can be time lags both for reporting of claims, where the insurer is not notified of a loss until long after it has occurred, and for final development of claims, where payments continue long after the loss has been reported. Also, the amounts of claims are uncertain before they have fully developed. These factors contribute to the difficulty of the loss reserving problem, for which extensive literature exists and active research is being done. We refer the reader to \citet{england2002stochastic} for a survey of the problem and existing techniques.

Deep learning has garnered increasing interest in recent years due to successful applications in many fields \citep{lecun2015deep} and has recently made its way into the loss reserving literature. \citet{wuthrich2018neural} augments the traditional chain ladder method with neural networks to incorporate claims features, \citet{gabrielli2018individual} utilize neural networks to syntheisze claims data, and \citet{gabrielli2018neural} and \citet{gabrielli2019neural} embed classical parametric loss reserving models into neural networks. More specifically, the development in \citet{gabrielli2018neural} and \citet{gabrielli2019neural} proposes initializing a neural network so that, before training, it corresponds exactly to a classical model, such as the over-dispersed Poisson model. The training iterations then adjust the weights of the neural network to minimize the prediction errors, which can be interpreted as a boosting procedure.

In developing our framework, which we call DeepTriangle\footnote{A portmanteau of \emph{deep learning} and \emph{loss development triangle}.}, we also draw inspiration from the existing stochastic reserving literature. Works that propose utilizing data in addition to paid losses include \citet{quarg2004munich}, which uses incurred losses, and \citet{miranda2012double}, which incorporates claim count information. Moving beyond a single homogeneous portfolio, \citet{avanzi2016stochastic} considers the dependencies among lines of business within an insurer's portfolio, while \citet{peremans2018robust} proposes a robust general multivariate chain ladder approach to accommodate outliers. There is also a category of models, referred to as state space or adaptive models, that allow parameters to evolve recursively in time as more data is observed \citep{chukhrova2017state}. This iterative updating mechanism is similar in spirit to the continuous updating of neural network weights during model deployment.

The approach that we develop differs from existing works in many ways, and has the following advantages. First, it enables joint modeling of paid losses and claims outstanding for multiple companies simultaneously in a single model. In fact, the architecture can also accommodate arbitrary additional inputs, such as claim count data and economic indicators, should they be available to the modeler. Second, it requires no manual input during model updates or forecasting, which means that predictions can be generated more frequently than traditional processes, and, in turn, allows management to react to changes in the portfolio sooner.

The rest of the paper is organized as follows: Section \ref{prelim} provides a brief overview of neural network terminology, Section \ref{data-arch} discusses the dataset used and introduces the proposed neural network architecture, Section \ref{exps} defines the performance metrics we use to benchmark our models and discuss the results, and Section \ref{conclusion} concludes.

\hypertarget{prelim}{%
\section{Neural Network Preliminaries}\label{prelim}}

For comprehensive treatments of neural network mechanics and implementation, we refer the reader to \citet{goodfellow2016deep} and \citet{chollet2018deep}. A more actuarially oriented discussion can be found in \citet{wuthrich2019data}. In order to establish common terminology used in this paper, we present a brief overview in this section.

\begin{figure}

{\centering \includegraphics[width=0.8\linewidth]{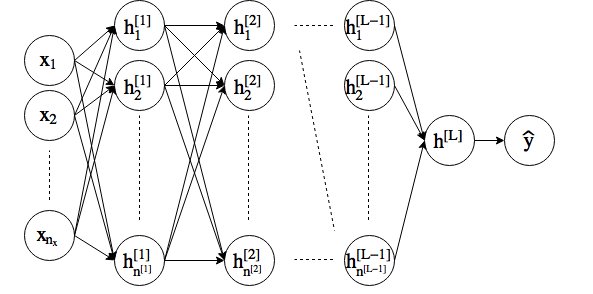} 

}

\caption{Feedforward neural network.}\label{fig:feedforward}
\end{figure}

We motivate the discussion by considering an example feedforward network with fully connected layers represented in Figure \ref{fig:feedforward}, where the goal is to predict an output \(y\) from input \(x\). The intermediate values, known as hidden layers and represented by \(h_j^{[l]}\), try to transform the input data into representations that successively become more useful at predicting the output. The nodes in the figure are computed, for each layer \(l = 1, \dots, L\), as

\begin{equation}
h_j^{[l]} = g^{[l]}(z_j^{[l]}),
\end{equation}

where

\begin{equation}
z_j^{[l]} = w_j^{[l]T}h^{[l-1]}+ b_j^{[l]},
\end{equation}

for \(j = 1, \dots, n^{[l]}\). In these equations, a superscript \([l]\) denotes association with the layer \(l\), a subscript \(j\) denotes association with the \(j\)-th component of the layer, of which there are \(n^{[l]}\). The \(g^{[l]}\) (\(l = 1, \dots, L\)) are called activation functions, whose values \(h^{[l]}\) are known as activations. The vectors \(w_j^{[l]}\) and scalars \(b_j^{[l]}\) are known as weights and biases, respectively, and together represent the parameters of the neural network, which are learned during training.

For \(l = 1\), we define the previous layer activations as the input, so that the calculation for the first hidden layer becomes
\begin{equation}
h_j^{[1]} = g^{[1]}(w_j^{[1]T}x + b_j^{[1]}).
\end{equation}
Also, for the output layer \(l = L\), we compute the prediction
\begin{equation}
\hat{y} = h_j^{[L]} = g^{[L]}(w_{j}^{[L]T}h^{[L-1]} + b_j^{[L]}).
\end{equation}

We can then think of a neural network as a sequence of function compositions \(f = f_L \circ f_{L-1} \circ \dots \circ f_1\) parameterized as \(f(x; W^{[1]}, b^{[1]}, \dots, W^{[L]}, b^{[L]})\). Here, it should be mentioned that the \(g^{[l]}\) (\(l = 1, \dots, L\)) are chosen to be nonlinear, except for possibly in the output layer. These nonlinearities are key to the success of neural networks, because otherwise we would have a trivial composition of linear models.

Each neural network model is specified with a specific loss function, which is used to measure how close the model predictions are to the actual values. During model training, the parameters discussed above are iteratively updated in order to minimize the loss function. Each update of the parameters typically involves only a subset, or mini-batch, of the training data, and one complete pass through the training data, which includes many updates, is known as an epoch. Training a neural network often requires many passes through the data.

\hypertarget{data-arch}{%
\section{Data and Model Architecture}\label{data-arch}}

In this section, we discuss the dataset used for our experiments and the proposed model architecture.

\hypertarget{data-source}{%
\subsection{Data Source}\label{data-source}}

We use the National Association of Insurance Commissioners (NAIC) Schedule P triangles \citep{meyers2011loss}. The dataset corresponds to claims from accident years 1988-1997, with development experience of 10 years for each accident year. In Schedule P data, the data is aggregated into accident year-development year records. The procedure for constructing the dataset is detailed in \citet{meyers2015stochastic}.

Following \citet{meyers2015stochastic}, we restrict ourselves to a subset of the data which covers four lines of business (commercial auto, private personal auto, workers' compensation, and other liability) and 50 companies in each line of business. This is done to facilitate comparison to existing results.

We use the following variables from the dataset in our study: line of business, company code, accident year, development lag, incurred loss, cumulative paid loss, and net earned premium. Claims outstanding, for the purpose of this study, is derived as incurred loss less cumulative paid loss. The company code is a categorical variable that denotes which insurer the records are associated with.

\hypertarget{trainingtesting-setup}{%
\subsection{Training/Testing Setup}\label{trainingtesting-setup}}

Let indices \(1 \leq i \leq I\) denote accident years and \(1 \leq j \leq J\) denote development years under consideration. Also, let \(\{P_{i,j}\}\) and \(\{OS_{i,j}\}\) denote the \emph{incremental} paid losses and the \emph{total} claims outstanding, or case reserves, respectively.

Then, at the end of calendar year \(I\), we have access to the observed data

\begin{equation}
\{P_{i,j}: i = 1, \dots, I; j = 1, \dots, I - i + 1\}
\end{equation}

and

\begin{equation}
\{OS_{i,j}: i = 1, \dots, I; j = 1, \dots, I - i + 1\}.
\end{equation}

Assume that we are interested in development through the \(I\)th development year; in other words, we only forecast through the eldest maturity in the available data. The goal then is to obtain predictions for future values \(\{\widehat{P}_{i,j}: i = 2, \dots, I; j = i+1, \dots, I\}\) and \(\{\widehat{OS}_{i,j}: i = 2, \dots, I; j = i+1, \dots, I\}\). We can then determine ultimate losses (UL) for each accident year \(i = 1, \dots, I\) by calculating

\begin{equation}
\widehat{UL}_i = \left(\sum_{j = 1}^{I - i + 1} P_{i,j}\right) + \left(\sum_{j = I - i + 2}^I \widehat{P}_{i,j}\right).
\end{equation}

In our case, data as of year end 1997 is used for training. We then evaluate predictive performance on the development year 10 cumulative paid losses.

\hypertarget{response-and-predictor-variables}{%
\subsection{Response and Predictor Variables}\label{response-and-predictor-variables}}

In DeepTriangle, each training sample is associated with an accident year-development year pair, which we refer to thereinafter as a \emph{cell}. The response for the sample associated with accident year \(i\) and development year \(j\) is the sequence

\begin{equation}
(Y_{i,j},Y_{i,j+1},\dots,Y_{i,I - i + 1}), 
\end{equation}
where each \(Y_{i,j} = (P_{i,j} / NPE_{i}, OS_{i,j} / NPE_{i})\), and \(NPE_{i}\) denotes the net earned premium for accident year \(i\). Working with loss ratios makes training more tractable by normalizing values into a similar scale.

The predictor for the sample contains two components. The first component is the observed history as of the end of the calendar year associated with the cell:

\begin{equation}
(Y_{i,1}, Y_{i,2}, \dots, Y_{i,j-1}).
\end{equation}
In other words, for each accident year and at each evaluation date for which we have data, we attempt to predict future development of the accident year's paid losses and claims outstanding based on the observed history as of that date. While we are ultimately interested in \(P_{i,j}\), the paid losses, we include claims outstanding as an auxiliary output of the model. We elaborate on the reasoning behind this approach in the next section.

The second component of the predictor is the company identifier associated with the experience. Because we include experience from multiple companies in each training iteration, we need a way to differentiate the data from different companies. We discuss handling of the company identifier in more detail in the next section.

\hypertarget{model-architecture}{%
\subsection{Model Architecture}\label{model-architecture}}

\begin{figure}

{\centering \includegraphics[width=1\linewidth]{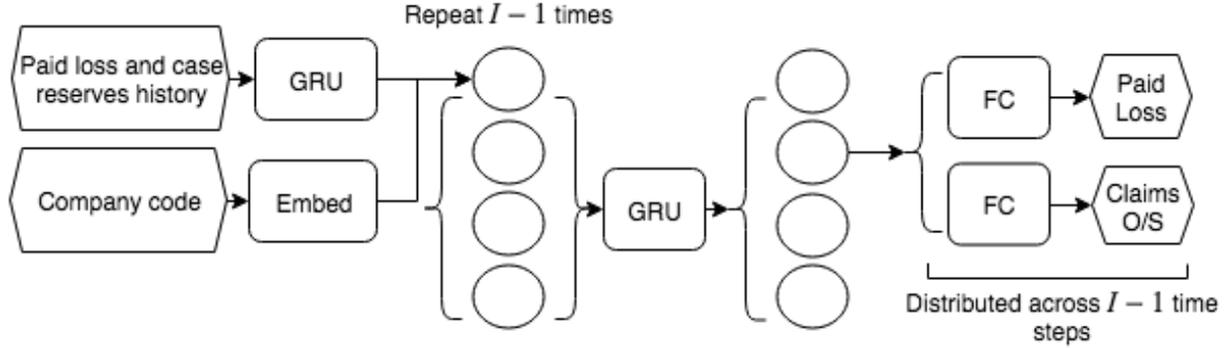} 

}

\caption{DeepTriangle architecture. \textit{Embed} denotes embedding layer, \textit{GRU} denotes gated recurrent unit, \textit{FC} denotes fully connected layer.}\label{fig:dt}
\end{figure}

As shown in Figure \ref{fig:dt}, DeepTriangle is a multi-task network \citep{caruana1997multitask} utilizing a sequence-to-sequence architecture \citep{sutskever2014sequence, DBLP:journals/corr/SrivastavaMS15} with two prediction goals: paid loss and claims outstanding. We construct one model for each line of business and each model is trained on data from multiple companies.

\hypertarget{multi-task-learning}{%
\subsubsection{Multi-Task Learning}\label{multi-task-learning}}

Since the two target quantities, paid loss and claims outstanding, are related, we expect to obtain better performance by jointly training than predicting each quantity independently. While \citet{caruana1997multitask} contains detailed discourse on the specific mechanisms of multi-task learning, we provide some heuristics on why it may improve predictions: by utilizing the reponse data for claims outstanding, we are effectively increasing the training data size since we are providing more signals to the learning algorithm; there may be hidden features, useful for predicting paid losses, that are more easily learned by trying to predict claims outstanding; also, by trying to predict claims outstanding during training, we are imposing a bias towards neural network weight configurations which perform that task well, which lessens the likelihood of arriving at a model that overfits to random noise.

\hypertarget{sequential-input-processing}{%
\subsubsection{Sequential Input Processing}\label{sequential-input-processing}}

\begin{figure}[h]

{\centering \includegraphics[width=0.7\linewidth]{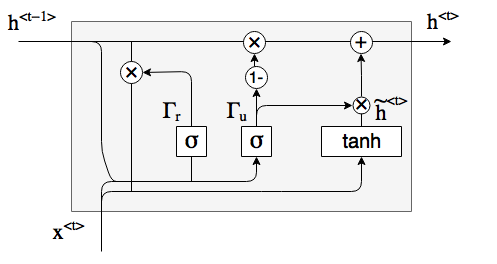} 

}

\caption{Gated recurrent unit.}\label{fig:gru}
\end{figure}

For handling the time series of paid losses and claims outstanding, we utilize gated recurrent units (GRU) \citep{chung2014empirical}, which is a type of recurrent neural network (RNN) building block that is appropriate for sequential data. A graphical representation of a GRU is shown in Figure \ref{fig:gru}, and the associated equations are as follows\footnote{Note the use of angle brackets to index position in a sequence rather than layers in a feedforward neural network as in Section \ref{prelim}.}:

\begin{equation}
\tilde{h}^{<t>} = \tanh(W_h[\Gamma_r h^{<t-1>}, x^{<t>}] + b_h)
\end{equation}
\begin{equation}
\Gamma_r^{<t>} = \sigma(W_r[h^{<t-1>}, x^{<t>}] + b_r)
\end{equation}
\begin{equation}
\Gamma_u^{<t>} = \sigma(W_u[h^{<t-1>},x^{<t>}] + b_u)
\end{equation}
\begin{equation}
h^{<t>} = \Gamma_u^{<t>} \tilde{h}^{<t>} + (1 - \Gamma_u^{<t>})h^{<t-1>}.
\end{equation}

Here, \(h^{<t>}\) and \(x^{<t>}\) represent the activation and input values, respectively, at time \(t\), and \(\sigma\) denotes the logistic sigmoid function defined as

\begin{equation}
\sigma(x) = \frac{1}{1 + \exp(-x)}\label{eq:eq1}.
\end{equation}
\(W_h\), \(W_r\), \(W_u\), \(b_h\), \(b_r\), and \(b_u\) are the appropriately sized weight matrices and biases to be learned. Intuitively, the activations \(h^{<t>}\) provide a way for the network to maintain state and ``remember'' values from early values of the input sequence. The values \(\tilde{h}^{<t>}\) can be thought of as candidates to replace the current state, and \(\Gamma_u^{<t>}\) determines the weighting between the previous state and the candidate state. We remark that although the GRU (and RNN in general) may seem opaque at first, they contain sequential instructions for updating weights just like vanilla feedforward neural networks (and can in fact be interpreted as such \citep{goodfellow2016deep}).

We first encode the sequential predictor with a GRU to obtain a summary encoding of the historical values. We then repeat the output \(I-1\) times before passing them to a decoder GRU that outputs its hidden state for each time step. The factor \(I-1\) is chosen here because for the \(I\)th accident year, we need to forecast \(I-1\) timesteps into the future. For both the encoder and decoder GRU modules, we use 128 hidden units and a dropout rate of 0.2. Here, dropout refers to the regime where, during training, at each iteration, we randomly set the output of the hidden units to zero with a specified probability, in order to reduce overfitting \citep{srivastava2014dropout}. Intuitively, dropout accomplishes this by approximating an ensemble of sub-networks that can be constructed by removing some hidden units.

\hypertarget{company-code-embeddings}{%
\subsubsection{Company Code Embeddings}\label{company-code-embeddings}}

The company code input is first passed to an embedding layer. In this process, each company is mapped to a fixed length vector in \(\mathbb{R}^k\), where \(k\) is a hyperparameter. In our case, we choose \(k = \text{number of levels} - 1 = 49\), as recommended in \citet{DBLP:journals/corr/GuoB16}. In other words, each company is represented by a vector in \(\mathbb{R}^{49}\). This mapping mechanism is part of the neural network and hence is learned during the training of the network, instead of in a separate data preprocessing step, so the learned numerical representations are optimized for predicted the future paid losses. Companies that are similar in the context of our claims forecasting problem are mapped to vectors that are close to each other in terms of Euclidean distance. Intuitively, one can think of this representation as a proxy for characteristics of the companies, such as size of book and case reserving philosophy. Categorical embedding is a common technique in deep learning that has been successfully applied to recommendation systems \citep{Cheng_2016} and retail sales prediction \citep{DBLP:journals/corr/GuoB16}. In the actuarial science literature, \citet{richman2018neural} utilize embedding layers to capture characteristics of regions in mortality forecasting, while \citet{gabrielli2018neural} apply them to lines of business factors in loss reserving.

\hypertarget{fully-connected-layers-and-outputs}{%
\subsubsection{Fully Connected Layers and Outputs}\label{fully-connected-layers-and-outputs}}

Each timestep of the decoded sequence from the GRU decoder is then concatenated with the company embedding output. The concatenated values are then passed to two subnetworks of fully connected layers, each of which shares weights across the timesteps. The two subnetworks correspond to the paid loss and case outstanding predictions, respectively, and each consists of a hidden layer of 64 units with a dropout rate of 0.2, followed by an output layer of 1 unit to represent the paid loss or claims outstanding at a time step.

Rectified linear unit (ReLU) \citep{nair2010rectified}, defined as

\begin{equation}
x \mapsto \max(0, x),
\end{equation}

is used as the activation function (which we denote by \(g\) in Section \ref{prelim}) for all fully connected layers, including both of the output layers. We remark that this choice of output activation implies we only predict nonnegative cash flows, i.e.~no recoveries. This assumption is reasonable for the dataset we use in our experiments, but may be modified to accommodate other use cases.

\hypertarget{deployment-considerations}{%
\subsection{Deployment Considerations}\label{deployment-considerations}}

While one may not have access to the latest experience data of competitors, the company code predictor can be utilized to incorporate data from companies within a group insurer. During training, the relationships among the companies are inferred based on historical development behavior. This approach provides an automated and objective alternative to manually aggregating, or clustering, the data based on knowledge of the degree of homogeneity among the companies.

If new companies join the portfolio, or if the companies and associated claims are reorganized, one would modify the embedding input size to accommodate the new codes, leaving the rest of the architecture unchanged, then refit the model. The network would then assign embedding vectors to the new companies.

Since the model outputs predictions for each triangle cell, one can calculate the traditional age-to-age, or loss development, factors (LDF) using the model forecasts. Having a familiar output may enable easier integration of DeepTriangle into existing actuarial workflows.

Insurers often have access to richer information than is available in regulatory filings, which underlies the experiments in this paper. For example, in addition to paid and incurred losses, one may include claim count triangles so that the model can also learn from, and predict, frequency information.

\hypertarget{exps}{%
\section{Experiments}\label{exps}}

We now describe the performance metrics for benchmarking the models and training details, then discuss the results.

\hypertarget{evaluation-metrics}{%
\subsection{Evaluation Metrics}\label{evaluation-metrics}}

We aim to produce scalar metrics to evaluate the performance of the model on each line of business. To this end, for each company and each line of business, we calculate the actual and predicted ultimate losses as of development year 10, for all accident years combined, then compute the root mean squared percentage error (RMSPE) and mean absolute percentage error (MAPE) over companies in each line of business. Percentage errors are used in order to have unit-free measures for comparing across companies with vastly different sizes of portfolios. Formally, if \(\mathcal{C}_l\) is the set of companies in line of business \(l\),

\begin{equation}
MAPE_l = \frac{1}{|\mathcal{C}_l|}\sum_{C\in\mathcal{C}_l}\left|\frac{\widehat{UL}_C - UL_C}{UL_C}\right|,
\end{equation}

and

\begin{equation}
RMSPE_l = \sqrt{\frac{1}{|\mathcal{C}_l|}\sum_{C\in\mathcal{C}_l}\left(\frac{\widehat{UL}_C - UL_C)}{UL_C}\right)^2}
\end{equation}

where \(\widehat{UL}_C\) and \(UL_C\) are the predicted and actual cumulative ultimate losses, respectively, for company \(C\).

An alternative approach for evaluation could involve weighting the company results by the associated earned premium or using dollar amounts. However, due to the distribution of company sizes in the dataset, the weights would concentrate on a handful of companies. Hence, to obtain a more balanced evaluation, we choose to report the unweighted percentage-based measures outlined above. We note that the evaluation of reserving models is an ongoing area of research; and refer the reader to \citet{martinek2019analysis} for a recent analysis.

\hypertarget{implementation-and-training}{%
\subsection{Implementation and Training}\label{implementation-and-training}}

The loss function is computed as the average over the forecasted time steps of the mean squared error of the predictions. The losses for the outputs are then averaged to obtain the network loss. Formally, for the sample associated with cell \((i, j)\), we can write the per-sample loss as

\begin{equation}
\frac{1}{I-i+1-(j-1)}\sum_{k = j}^{I-i+1}\frac{(\widehat{P_{i,k}} - P_{i,k})^2 + (\widehat{OS_{i,k}} - OS_{i,k})^2}{2}.
\end{equation}

For optimization, we use the \textsc{AMSGrad} \citep{j.2018on} variant of \textsc{adam} with a learning rate of 0.0005. We train each neural network for a maximum of 1000 epochs with the following early stopping scheme: if the loss on the validation set does not improve over a 200-epoch window, we terminate training and revert back to the weights on the epoch with the lowest validation loss. The validation set used in the early stopping criterion is defined to be the subset of the training data that becomes available after calendar year 1995. For each line of business, we create an ensemble of 100 models, each trained with the same architecture but different random weight initialization. This is done to reduce the variance inherent in the randomness associated with neural networks.

We implement DeepTriangle using the keras R package \citep{chollet2017kerasR} and TensorFlow \citep{tensorflow2015-whitepaper}, which are open source software for developing neural network models. Code for producing the experiment results is available online.\footnote{\url{https://github.com/kasaai/deeptriangle}.}

\hypertarget{results-and-discussion}{%
\subsection{Results and Discussion}\label{results-and-discussion}}

\begin{table}[t]

\caption{\label{tab:unnamed-chunk-1}\label{tab:table1}Performance comparison of various models. DeepTriangle and AutoML are abbreviated do DT and ML, respectively.}
\centering
\begin{tabular}{lrrrrr>{\bfseries}r}
\toprule
Line of Business & Mack & ODP & CIT & LIT & ML & DT\\
\midrule
\addlinespace[0.3em]
\multicolumn{7}{l}{\textbf{MAPE}}\\
\hspace{1em}Commercial Auto & 0.060 & 0.217 & 0.052 & 0.052 & 0.068 & 0.043\\
\hspace{1em}Other Liability & 0.134 & 0.223 & 0.165 & 0.152 & 0.142 & 0.109\\
\hspace{1em}Private Passenger Auto & 0.038 & 0.039 & 0.038 & 0.040 & 0.036 & 0.025\\
\hspace{1em}Workers' Compensation & 0.053 & 0.105 & 0.054 & 0.054 & 0.067 & 0.046\\
\addlinespace[0.3em]
\multicolumn{7}{l}{\textbf{RMSPE}}\\
\hspace{1em}Commercial Auto & 0.080 & 0.822 & 0.076 & 0.074 & 0.096 & 0.057\\
\hspace{1em}Other Liability & 0.202 & 0.477 & 0.220 & 0.209 & 0.181 & 0.150\\
\hspace{1em}Private Passenger Auto & 0.061 & 0.063 & 0.057 & 0.060 & 0.059 & 0.039\\
\hspace{1em}Workers' Compensation & 0.079 & 0.368 & 0.080 & 0.080 & 0.099 & 0.067\\
\bottomrule
\end{tabular}
\end{table}

In Table \ref{tab:table1} we tabulate the out-of-time performance of DeepTriangle against other models: the Mack chain-ladder model \citep{mack1993distribution}, the bootstrap ODP model \citep{england2002stochastic}, an AutoML model, and a selection of Bayesian Markov chain Monte Carlo (MCMC) models from \citet{meyers2015stochastic} including the correlated incremental trend (CIT) and leveled incremental trend (LIT) models. For the stochastic models, we use the means of the predictive distributions as the point estimates to which we compare the actual outcomes. For DeepTriangle, we report the averaged predictions from the ensembles.

The AutoML model is developed by automatically searching over a set of common machine learning techniques. In the implementation we use, it trains and cross-validates a random forest, an extremely-randomized forest, a random grid of gradient boosting machines, a random grid of deep feedforward neural networks, and stacked ensembles thereof \citep{h2o_R_package}. Details of these algorithms can be found in \citet{friedman2001elements}. Because the machine learning techniques produce scalar outputs, we use an iterative forecasting scheme where the prediction for a timestep is used in the predictor for the next timestep.

We see that DeepTriangle improves on the performance of the popular chain ladder and ODP models, common machine learning models, and Bayesian stochastic models.

\begin{figure}

{\centering \includegraphics[width=1\linewidth]{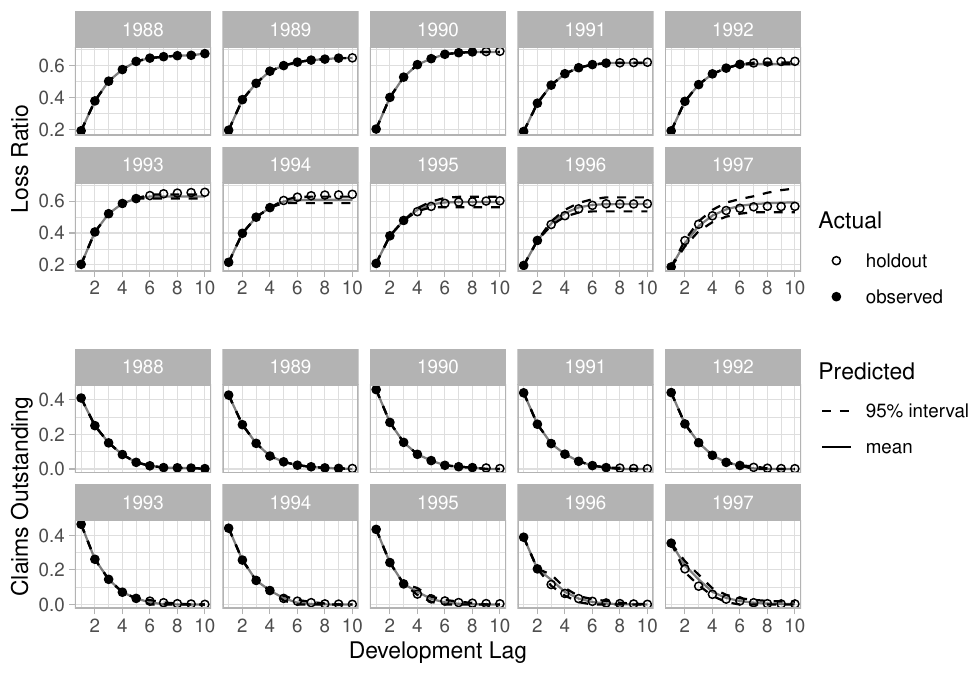} 

}

\caption{Development by accident year for Company 1767, commercial auto.}\label{fig:fig3}
\end{figure}

\begin{figure}

{\centering \includegraphics[width=1\linewidth]{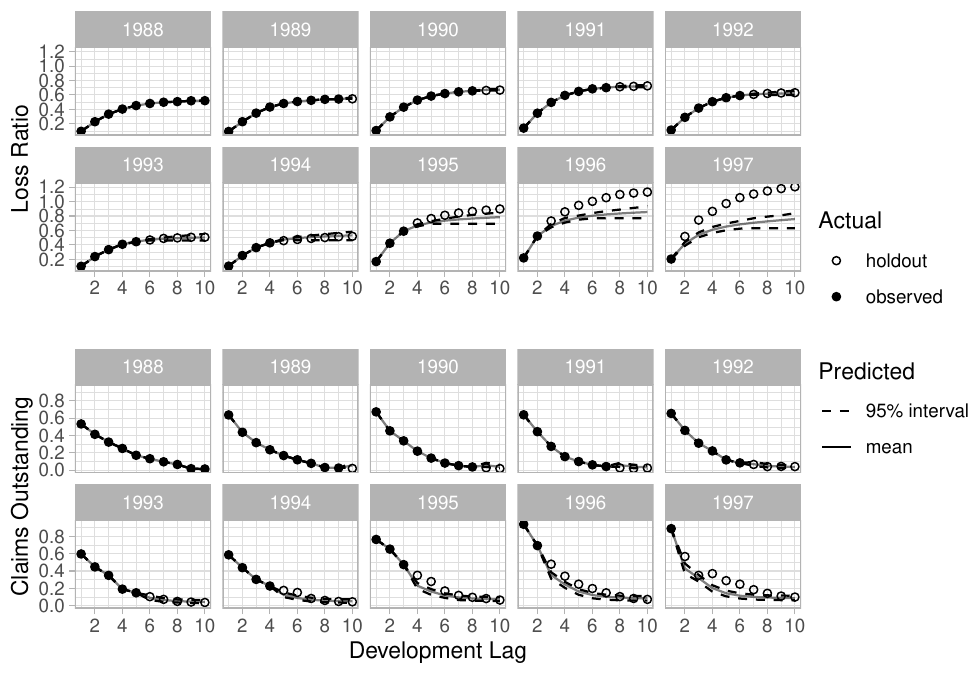} 

}

\caption{Development by accident year for Company 337, workers' compensation.}\label{fig:fig4}
\end{figure}

In addition to aggregated results for all companies, we also investigate qualitatively the ability of DeepTriangle to learn development patterns of individual companies. Figures \ref{fig:fig3} and \ref{fig:fig4} show the paid loss development and claims outstanding development for the commercial auto line of Company 1767 and the workers' compensation line of Company 337, respectively. We see that the model captures the development patterns for Company 1767 reasonably well. However, it is unsuccessful in forecasting the deteriorating loss ratios for Company 337's workers' compensation book.

We do not study uncertainty estimates in this paper nor interpret the forecasts as posterior predictive distributions; rather, they are included to reflect the stochastic nature of optimizing neural networks. We note that others have exploited randomness in weight initialization in producing predictive distributions \citep{NIPS2017_7219}, and further research could study the applicability of these techniques to reserve variability.

\hypertarget{conclusion}{%
\section{Conclusion}\label{conclusion}}

We introduce DeepTriangle, a deep learning framework for forecasting paid losses. Our models are able to attain performance comparable, by our metrics, to modern stochastic reserving techniques, without expert input. This means that one can automate model updating and report production at the desired frequency (although we note that, as with any automated machine learning system, a process involving expert review should be implemented). By utilizing neural networks, we can incorporate multiple heterogeneous inputs and train on multiple objectives simultaneously, and also allow customization of models based on available data. To summarize, this framework maintains accuracy while providing automatability and extensibility.

We analyze an aggregated dataset with limited features in this paper because it is publicly available and well studied, but one can extend DeepTriangle to incorporate additional data, such as claim counts.

Deep neural networks can be designed to extend recent efforts, such as \citet{wuthrich2018machine}, on applying machine learning to claims level reserving. They can also be designed to incorporate additional features that are not handled well by traditional machine learning algorithms, such as claims adjusters' notes from free text fields and images.

While this study focuses on prediction of point estimates, future extensions may include outputting distributions in order to address reserve variability.

%

\vspace{6pt}

\acknowledgments{We thank Sigrid Keydana, Ronald Richman, the anonymous reviewers, and the volunteers on the Casualty Actuarial Society Committee on Reserves (CASCOR) who helped to improve the paper through helpful comments and discussions.}


\conflictsofinterest{The author declares no conflict of interest.}





\reftitle{References}
\externalbibliography{yes}
\bibliography{manuscript.bib}



\end{document}